# Extending the optical absorption in a lumped element meander structure to far-infrared wavelengths

Shekhar Chandra Pandey, Shilpam Sharma, Anudeep Singh, Utkarsh Pandey, S. S. Prabhu, Bhaskar Biswas, Sona Chandran and M. K. Chattopadhyay

*Abstract—* Superconducting radiation detectors typically exhibit detection and single-photon sensitivity limited to the mid-infrared wavelength range. Extending their detection capabilities into the far-infrared range (>10 $\mu$m) requires careful selection of substrate materials and detector geometries. The overall detection efficiency of such detectors is inherently linked to their absorption and coupling efficiencies. In this study, the resonator geometry and its absorption efficiency were estimated using electromagnetic simulations in CST Microwave Studio simulation software for a lumped-element meander structure. Simulations were performed for 12– 50 $\mu$m wavelength range, which corresponds to the Infrared Free Electron Laser (IR-FEL) at RRCAT, Indore. The absorption in the meander inductor was significantly influenced by the substrate material and thickness, as well as the impedance matching between the detector structure and the incident photon medium. The results indicate that $SiO_2$ and diamond substrates are well-suited for developing lumped element kinetic inductance detectors (LEKID) in the above wavelength range. Optimised meander geometries on diamond substrates demonstrated absorption efficiencies of up to 95% for narrow bandwidths and greater than 50% for wide bandwidths. Based on the simulations, a 30-pixel LEKID structure was fabricated using electron beam lithography on a 500 $\mu$m $SiO_2$-coated Si substrate. The resonator was made with a 20 nm thick $Ti_{40}V_{60}$ alloy thin film. Experimental determination of absorption efficiency was also performed through simultaneous transmission and reflection measurements of the 30-pixel LEKID. The results indicate that in the 14– 26 $\mu$m wavelength range of the IR-FEL light, the LEKID achieved an absorption efficiency of up to 75% for certain wavelengths. Combined simulation and experimental studies demonstrate that the designed LEKID structure is well-suited for detecting far-infrared wavelengths above 10 μm, offering high absorption efficiency.

*Index Terms—* Absorption, CST simulation, Electromagnetic simulation, Free electron laser, Lumped element kinetic inductance detector, Superconducting thin films, Titanium-Vanadium alloy.

## I. INTRODUCTION

TRADITIONAL radiation detectors using semiconducting materials as the detector element have limitations in detecting long wavelengths, where the photon energies are lower. In these detectors, the band gap of the semiconductor materials defines the range of photon energies that can be detected. In addition, the thermal noise resulting from their high operating temperatures also limits their use at the long wavelengths (e.g., far-IR) [1-3]. If the photon energy is lower than the band gap of the material, the photons cannot excite electrons from the valence band to the conduction band. The semiconductor-based IR detectors routinely used in the FTIR spectrometers typically use materials like InSb and HgCdTe (MCT). However, InSb-based detectors are sensitive only up to about 5.5 $\mu$m, and the $Hg_{1-x}Cd_xTe$ alloy-based detectors are limited to 17 – 20 $\mu$m.

Superconductors, on the other hand, have much smaller energy gaps and can easily detect radiations up to mid-IR wavelengths with excellent sensitivity and efficiency [3]. Carefully designed detectors using superconducting materials with even smaller energy gaps can be utilised for ultra-sensitive radiation detection in the far-IR bands of the spectrum [4]. Based on the response of the superconductors to the electromagnetic radiation, several different detection schemes are employed in the superconducting radiation detectors. The key types include kinetic inductance detectors (KIDs), superconducting nanowire single-photon detectors (SNSPDs), superconducting tunnel junctions (STJs), and transition edge sensors (TESs). These detectors exhibit excellent single-photon sensitivity across a broad wavelength range varying from X-

(*Corresponding author: shilpam@rrcat.gov.in*).
Shekhar Chandra Pandey, Sona Chandran and M. K. Chattopadhyay are with the Free Electron Laser & Utilization Section, Raja Ramanna Center for Advanced Technology, Indore, Madhya Pradesh, 452013, India and Homi Bhabha National Institute, Training School Complex, Anushakti Nagar, Mumbai, 400094, India. (email: shekharpandey7579@gmail.com, sona@rrcat.gov.in, maulindu@rrcat.gov.in )

Shilpam Sharma, Anudeep Singh and Bhaskar Biswas are with the Free Electron Laser & Utilization Section, Raja Ramanna Center for Advanced Technology, Indore, Madhya Pradesh, 452013, India. (email: shilpam@rrcat.gov.in, anudeep@rrcat.gov.in, bbiswas@rrcat.gov.in )

Utkarsh Pandey and S. S. Prabhu are with the Department of Condensed Matter Physics and Materials Science, Tata Institute of Fundamental Research, Mumbai, 400005, India. (email: utkarshkumarpandey790@gmail.com, prabhu@tifr.res.in ) rays to millimeter waves, with ultrafast electronic response times down to a few picoseconds [3-8].

The present work focuses on the development of lumped element KIDs (LEKID) for the 12 - 50 $\mu m$ band of the mid and far-IR range of EM spectrum, which, except for a couple of reports, remains largely unexplored. Very recently Day *et al.* showed single photon sensitivity of the kinetic inductance detector at 25 $\mu m$ [4]. However, the development of highly sensitive, broadband detectors in this wavelength range remains an open area of research. Researchers are increasingly utilizing this wavelength range to gain insights into several astronomical phenomena [5, 9-13]. In addition, molecular vibrational spectroscopic studies for the structural determination of unstable species, chemical kinetics, reaction dynamics, etc., require sensitive detectors in the Far-IR range [3].

Here, we report on the development of LEKIDs for the IR-Free Electron Laser (IR-FEL) facility at RRCAT, Indore, India [14], that can be continuously tuned to provide coherent, pulsed monochromatic IR light in the 12– 50 $\mu m$ band.

## II. THEORY AND DESIGN

In a Kinetic Inductance Detector (KID), the photons having energy greater than the superconducting energy gap of an absorbing material break the Cooper pairs, leading to the generation of quasi-particles and a consequent alteration in the absorbing material's electrical properties [13]. The increased presence of quasi-particles significantly influences the kinetic inductance of the material [15]. This change in Cooper pair density can be measured by observing shifts in the structure's resonance frequency, amplitude, or phase. The LEKID is a type of KID composed of a meander structure and interdigitated capacitors, which serve as the inductive and capacitive components, respectively. These capacitive and inductive parts set the geometrical resonant frequency of the structure as $\omega = \frac{1}{\sqrt{L_{total}C}}$, where $L_{total}$ is the sum of geometric inductance ($L_g$) and kinetic inductance ($L_K$). Maintaining a constant current density across the entire inductive section ensures the uniform detection sensitivity of the LEKID structure. A probe signal is transmitted through the matched microstrip feed line, to which the LEKID is inductively coupled. To optimise coupling efficiency, it is essential to match the effective impedance of the meander with the impedance of the medium through which the photon travels (free space impedance for direct illumination or the impedance of the substrate for illumination through substrate).

The optical efficiency of the LEKID geometry depends on several parameters, such as the normal state resistivity of the superconducting material, substrate dielectric constant, width (w) of the meander lines and spacing (s) between the meander lines. The effective sheet impedance of the meander section can be expressed as [16]

$$Z_{LEKID} = R_{eff} + j\omega_{ph}L_{eff} \quad (1)$$

$R_{eff}$ is the effective sheet resistance, and $L_{eff}$ is the effective sheet inductance as seen by a photon of frequency $\omega_{ph}$. $R_{eff}$ and $L_{eff}$ are defined as,

$$R_{eff} = R_{sheet}\frac{p}{w} \quad (2)$$

$$L_{eff} = \frac{p}{2\pi c}ln\left(cosec\frac{\pi p}{\omega}\right)z_{fs} \quad (3)$$

Here, $p$ is the pitch ($w+s$) of the meander structure, $Z_{fs}$ is the free space impedance (considering photons coming from free space), and $c$ is the speed of light in vacuum. Therefore, to achieve impedance matching, the superconducting material parameters and the filling factor (ratio of $p$ to $w$) are critical. Considering the incidence of photons from the free space, the optimal absorption can be achieved if the effective impedance of the LEKID device is matched to the free space impedance (377 Ω). Equation 2 shows that adjusting the sheet resistance of the film, as well as the width and spacing of the meander structure, one can achieve the desired impedance. Equation 2 is valid only when, $s \ll w$ and $w < \lambda$, where $\lambda$ is the wavelength of the incoming photon [17]. In our previous work on the electrical characterisation of $Ti_{40}V_{60}$ alloy thin films it was reported that different electrical properties of the alloy thin films could be tuned by adjusting the deposition pressure and sputtering current [18]. It is thus possible to optimise the superconducting transition temperature ($T_C$) and sheet resistance ($Rs$) of the $Ti_{40}V_{60}$ thin film. So, by choosing the material with appropriate parameters and adjusting the $s$ and $w$ of the structure, one can match the effective impedance for optimal absorption. Absorption of the incident radiation on the detector active material (absorbing metal in the form of a meander structure) is an important factor for the detector's overall efficiency. Literature suggested that using a back short improves the absorption in the resonator [17]. Consequently, we performed simulations both for the bare meander structure and the meander with a back short. When the radiation impinges from the front of the meander geometry, the thin layer of an appropriate material deposited on the back of the substrate acts as a back short. In addition to thickness, the choice of the substrate material is a crucial factor, as IR-absorption in the substrate can reduce the detector's performance.

To meet the lumped element criteria, the meander line and its spacing should be less than the wavelength of the resonance signal. Additionally, the pitch of the meander design is kept below $\lambda/3$, where $\lambda$ is the wavelength of the incoming radiation [19]. Therefore, optimising the meander line width and spacing, along with the sheet resistance of the material, ensures impedance matching and enhances the absorption of incoming radiation in the detector's active material.

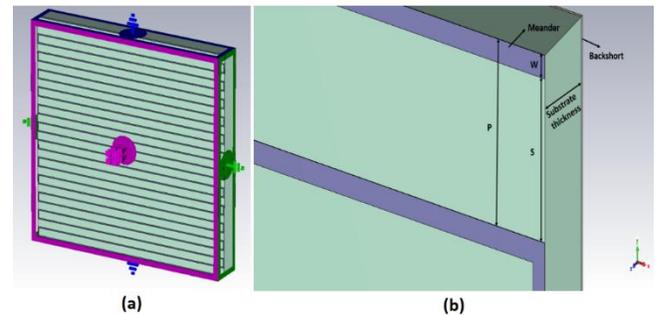

**Fig. 1. (a)** Schematic diagram of the meander structure on the substrate used for CST simulation. **(b)** Zoomed-in image of the schematic meander structure illustrating the meander and substrate parameters. *P*, *W*, and *S* represent the pitch, width, and separation of the meander lines.

## III. CST SIMULATIONS: RESULTS AND DISCUSSION

Electromagnetic simulations have been conducted to optimise the detector design for the 12- 50 $\mu$m wavelength range. The model design of the detector's radiation-sensitive part (meander) is presented in Fig. 1. The substrate material and its thickness (for the back-short) were optimised through CST simulations for efficient absorption in the above mentioned wavelength range. The boundary conditions for the simulation are shown in the Fig. 1(a). In this figure, the left and right box walls operate like a perfect electric conductor, where all the tangential electric fields and normal magnetic fluxes are set to zero. Meanwhile, the upper and lower box walls operate like a perfect magnetic conductor, where all the tangential magnetic fields and normal electric fluxes are set to zero. Ports on both the ends of the box (normal to the absorbing meander area) are set to open space. The simulated absorption in $SiO_2$ bare substrate and $SiO_2$ substrate with various back short distances (starting from t = 1 micron) are presented in Fig. 2(a) and (b) respectively. Additionally, beyond 25 $\mu$m, $SiO_2$ exhibitted minimal absorption, which makes it an ideal substrate material for that range. Hereafter we have selected the $SiO_2$ substrate with 2.2 $\mu$m thickness and optimised the absorption in the meander structure with this substrate thickness.

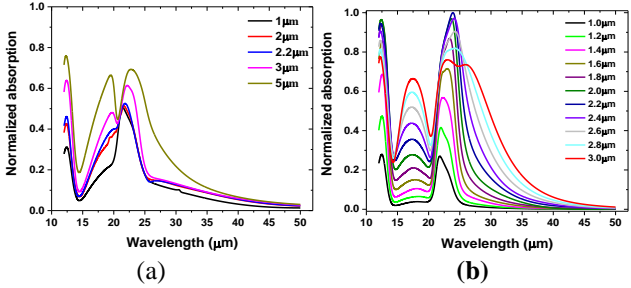

Fig. 2. (a) Absorption in the $SiO_2$ bare substrate having different thicknesses. (b) Absorption in the $SiO_2$ substrate with a back-short width varying from 1 to 3 $\mu$m.

Fig. 3(a) shows the absorption in the meander section with different sheet resistances starting from 10 $\Omega$/Sq., having a constant substrate thickness of 2.2 $\mu$m. The results show that a meander structure with a sheet resistance greater than 50 $\Omega$/Sq. can achieve absorption exceeding 50%. However, after 90 $\Omega$/Sq., there is a decline in absorption. To further optimise the line width (LW) and separation (LS) of the meander, we have fixed the sheet resistance value of 60 $\Omega$/Sq. Subsequently, the absorption in the meander section was simulated with different inductor line widths and separations, while keeping optimised substrate thickness and sheet resistance at t = 2.2 $\mu$m and $R_S$ = 60 $\Omega$/Sq respectively. Results demonstrated that a 0.5 $\mu$m LW and 1.5 $\mu$m LS meander structure achieves over 45% absorption beyond 25 $\mu$m (Fig. 3(b)). Adjusting the width and separation of the meander structure could yield 70- 80% absorption efficiency over a narrow bandwidth. The back short distance chosen to be an odd multiple of $\lambda_D/4$, where $\lambda_D = \lambda/\sqrt{\varepsilon_r}$ is the wavelength to be detected in the substrate and $\varepsilon_r$ is the dielectric constant of the substrate. The odd multiple of $\lambda_D/4$ ensures that the back short reflects the incoming radiation toward the detector so that the reflected signal adds constructively to the incident signal and enhances the resonance absorption.

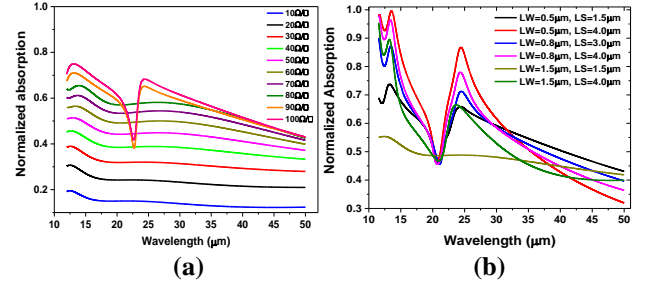

Fig. 3. (a) Absorption in the meander structure on a $SiO_2$ substrate having different sheet resistance from 10 to 100 $\Omega$/sq. (b) Absorption in the meander structure having different meander line widths and separation, while keeping fixed substrate thickness and the film's sheet resistance.

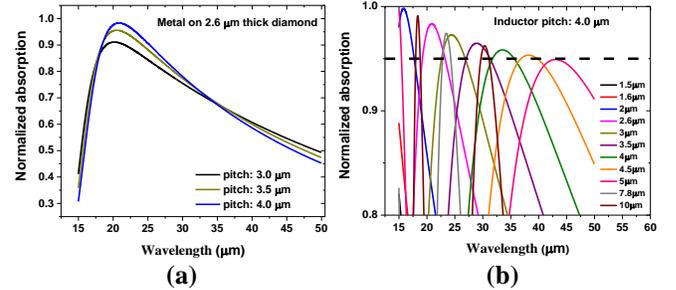

Fig. 4. (a) Absorption in the meander structure with different pitches on a 2.6 $\mu$m thick diamond substrate. Greater than 95% absorption is possible for a narrow bandwidth. (b). Absorption in the 4 $\mu$m pitch meander structure on a diamond substrate with different back short distances from 1.5 $\mu$m to 10 $\mu$m.

While $SiO_2$ shows excellent performance in the required wavelength range, considering the synthesis and mechanical handling challenges for the thicknesses needed for $SiO_2$, we have also done the simulation on a diamond substrate. Diamond exhibited excellent transmission (negligible absorption) for the 12- 50 $\mu$m wavelength range, hardness ≥10 on the Mohs scale (allowing for free-standing membranes of up to a few microns thickness), and well-established chemical vapour deposited (CVD) growth on Si substrates [20]. A simulation was performed on a diamond substrate with meander geometry having a different meander pitch. Fig. 4(a) shows that near-100% absorption may be achieved when the inductor pitch is 4 $\mu$m (LW = 0.5 $\mu$m and LS = 3.5 $\mu$m) with a 2.6 $\mu$m thick back short. This designed geometry effectively increases the sheet resistance of the film by a factor of 8. Subsequently, various substrate thicknesses (ranging from 1 to 10 microns) were explored. Fig. 4(b) shows that by adjusting the substrate thickness and hence the back short distance from the detector, a detection efficiency of ≥ 95% may be achieved for the entire wavelength range of 12– 50 $\mu$m. Notably, at a $\lambda_D/(4)$ backshort distance (equivalent to 2.6 $\mu$m for 25 $\mu$m wavelength), broadband absorption of ≥50% can be achieved. It was observed that broadband absorption may be achieved with a $\lambda_D/(4)$ back short distance, and the bandwidth selectivity becomes narrower with higher odd multiples of $\lambda_D/(4)$. This confirms that the back short distance (substrate thickness) determines the efficient detection bandwidth of the LEKID.

Fig. 5(a) presents the power density distribution within the meander structure. The power density variations across nine maps correspond to changes in the incoming photon wavelength. The simulated structure is designed for 25 $\mu$m wavelength. As shown in the Fig 5(a), the maximum absorption occurs at 25 $\mu$m, while densities are lower for wavelengths both below and above this value. The simulation results further confirm the polarisation sensitivity of the meander structure, as shown in Fig. 5(a). Fig. 5(b) also indicates the direction of the electric field vector of the incoming wave. The horizontal meander lines are aligned with the direction of this electric field vector, while the smaller vertical lines may be considered negligible or treated as transparent for the incoming wave.

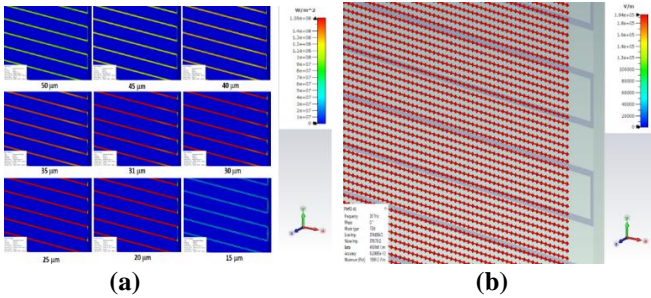

**Fig. 5. (a)** Power density maps of the 4 $\mu$m pitch inductor on a 2.6 $\mu$m thick diamond substrate, designed for a wavelength of 25 $\mu$m. At this wavelength, the maximum power density appears along the horizontal lines, while power density in the vertical lines is negligible. Above and below 25 $\mu$m, the power density in the meander lines decreases. **(b)** The electric field vector of the incident wave is aligned parallel to the horizontal wire. The red arrows indicate polarisation sensitivity of the meander structure, which is aligned with the horizontal meander lines.

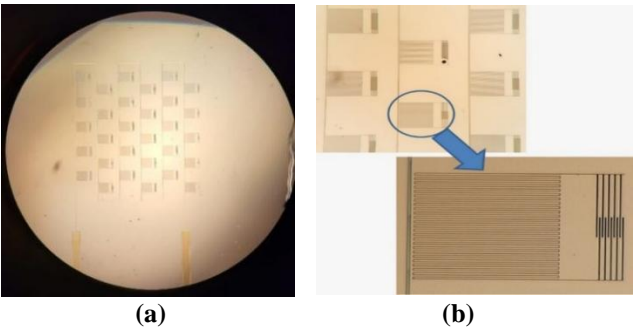

**Fig. 6. (a)** 30-pixel LEKID structure on a 500-thick SiO$_2$/Si substrate. **(b)** Enlarged image of a single pixel LEKID.

The theoretical design of the 30-pixel LEKID was modelled using the Phidl tool [21] in Python and Klayout (Layout Editor; KLayout 0.28.7), as illustrated in the Fig. 6(a). The inductor line width is 0.5 $\mu$m, with a separation of 3.5 $\mu$m. The high simulated absorption efficiency (> 95%) with both SiO$_2$ and diamond substrates suggests the potential for developing a sensitive LEKID for far-IR wavelength in the range 12- 50 $\mu$m.

## IV. FABRICATION AND ABSORPTION MEASUREMENT ON THE 30-PIXEL LEKID STRUCTURE

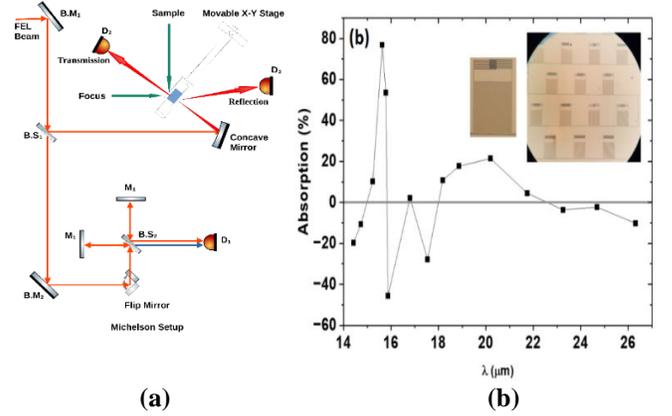

**Fig. 7. (a)** Schematic diagram of the absorption measurement using IR-FEL. **(b)** Experimentally measured absorption in a 30-pixel LEKID.

After simulating the absorption in the LEKID geometry, we initially fabricated the LEKID structure on a 500 $\mu$m thick SiO$_2$-coated Si substrate using electron beam lithography followed by reactive ion etching. Fig. 6(a) shows the fabricated 30-pixel LEKID structure, and Fig. 6(b) shows its enlarged image highlighting a single pixel. We performed simultaneous transmission and reflection measurements using our Infrared Free electron Laser (IR-FEL) for the experimental estimation of the absorption in the designed geometry. We recorded the corresponding intensities using WiredSense SPY pyroelectric detectors. The wavelength of the IR-FEL light was measured using a Michelson interferometer built in-house. Due to the near-continuous wavelength tunability of the IR-FEL, these measurements allowed us to calculate the wavelength-dependent absorption characteristics of the LEKID structure. Fig. 7(a) shows the schematic diagram of the absorption measurement set-up, and Fig. 7(b) shows the absorption of the 30-pixel structure at different wavelengths ranging from 14 $\mu$m to 26 $\mu$m. The figure also illustrates that at certain frequencies; the absorption is negative. This negative absorption could be due to surface plasmonic resonance, which is the collective oscillation of electrons bound to a metallic surface. This behaviour has been reported in the literature due to the periodic arrays of sub-wavelength apertures in thin metal films [22-24]. These structures are periodically patterned on the scale of the wavelength of light, and they may be used to manipulate and control the propagation parameters of electromagnetic radiation.

## V. CONCLUSION

In the present work, we have explored the potential of the lumped element kinetic inductance detector in the far-infrared wavelength range, specifically from 12- 50 $\mu m$. We have simulated and analysed the theoretical absorption response of a lumped element inductor structure in $SiO_2$ and diamond substrates. Our findings suggest that it is possible to achieve impedance-matched coupling within the resonator by optimising the film's meander geometry and sheet resistance. For the specified wavelength range, using a 4 $\mu m$ inductor pitch (0.5 $\mu m$ inductor width and 3.5 $\mu m$ inductor spacing), broadband absorption can be achieved with a backshort distance of $\lambda_D/4$, while the bandwidth selectivity narrows with higher odd multiples of $\lambda_D/4$. Our simulations indicate that LEKID detectors fabricated on $SiO_2$ and diamond substrates can operate effectively in the 12 - 50 $\mu m$ wavelength range, provided that the back short distance and resonator geometry are appropriately configured. We have found a high absorption (> 95%) in the diamond substrate across the IR-FEL wavelength. Additionally, we have experimentally measured the absorption in a 30-pixel LEKID geometry on a 500 $\mu m$ $SiO_2$/Si substrate.

ACKNOWLEDGMENT

We thank Mr. Rahul Gaur, ABPS, RRCAT, for his help in CST simulations and Mr. Ajinkya Punjal, TIFR Bombay, for his valuable discussions and support in electron beam lithography.